\documentclass[numberedappendix]{emulateapj}
\usepackage{amsmath}  

\providecommand*{\diff}{{\rm d}}
\providecommand*{\deriv}[3][]{\frac{\diff^{#1}#2}{\diff #3^{#1}}}
\providecommand{\abs}[1]{\left\vert#1\right\vert}

\newcommand{\cm}{{\rm cm}}
\newcommand{\GeV}{{\rm GeV}}
\newcommand{\eg}{{\it e.g.}}

\newcommand{\hyperlink}[1]{\anchor{#1}{\url{#1}}}

\newcommand{\vect}[1]{\mathbf{#1}}

\newcommand{\norm}{\mathcal{N}}
\newcommand{\btot}{B_\mathrm{tot}}
\newcommand{\cutoff}{E_\mathrm{cutoff}}
\newcommand{\Recur}{\mathcal{R}}
\newcommand{\recur}{\rho}

\newcommand{\ee}{{electron-electron}}
\newcommand{\ep}{{electron-proton}}
\newcommand{\ez}{{electron-ion}}
\newcommand{\srcut}{{\sc srcut}}
\newcommand{\sresc}{{\sc sresc}}
\newcommand{\xspec}{{\sc xspec}}
\newcommand{\pvm}{{\sc pvm}}
\newcommand{\slang}{{\rm S-Lang}}
\newcommand{\isis}{{\sc isis}}
\newcommand{\isisurl}{http://space.mit.edu/cxc/isis}

\slugcomment{Accepted for publication in ApJ Supplement}

\shorttitle{Nonthermal Spectra}
\shortauthors{Houck \& Allen}

\begin{document}
\title{Models for Nonthermal Photon Spectra}
\author{John C. Houck and Glenn E. Allen}
\affil{
MIT, Kavli Institute for Astrophysics and Space Research,
77 Massachusetts Avenue, Cambridge, MA 02139-4307
}
\email{houck@space.mit.edu}

\begin{abstract}
We describe models of nonthermal photon emission from a
homogeneous distribution of relativistic electrons and protons.
Contributions from the synchrotron, inverse Compton, nonthermal
bremsstrahlung and neutral-pion decay processes are computed
separately using a common parameterization of the underlying
distribution of nonthermal particles. The models are intended
for use in fitting spectra from multi-wavelength observations
and are designed to be accurate and efficient. Although our
applications have focused on Galactic supernova remnants, the
software is modular, making it straightforward to customize for
different applications. In particular, the shapes of the
particle distribution functions and the shape of the seed
photon spectrum used by the inverse Compton model are defined
in separate modules and may be customized for specific
applications. We assess the accuracy of these models by using a
recurrence relation and by comparing them with analytic results
and with previous numerical work by other authors.
\end{abstract}


\section{Introduction}

Cosmic-rays with energies up to about 1000\,TeV are thought to
be accelerated in supernova remnant shocks \citep{gaisser:94a}.
In the simplest, idealized model of the acceleration process,
test particles interact with a shock discontinuity to produce a
power-law momentum distribution of cosmic-rays.  Photon spectra
produced by power-law momentum spectra, as in the \xspec\
models \srcut\ and \sresc\ \citep{reynolds:98a,reynolds:99a},
have been widely used to fit the X-ray synchrotron
spectra of several SNRs.

However, in the diffusive shock acceleration picture
\citep{blandford:87a, bell:87a, ellison:91a, berezhko:99a},
nonlinear processes are expected to cause some deviation, or
curvature, away from a power-law momentum distribution. Recent
observations of Cas A \citep{jones:03a} and SN1006
\citep{allen:05a} have shown evidence for a curved synchrotron
spectrum. Such spectra can be studied using simulations of
diffusive shock acceleration in supernova remnants \citep[see
\eg][]{ellison:91a, baring:99a, bykov:00a}. These simulations
predict the momentum distribution of nonthermal particles and
the resulting photon emission spectrum. Unfortunately, such
detailed calculations are still too time-consuming for
widespread use in iterative fitting of models to observed
spectra.

Therefore, we have developed models of the synchrotron
(\S\ref{sec:sync}), inverse Compton (\S\ref{sec:invc}),
nonthermal bremsstrahlung (\S\ref{sec:ntbrem}) and neutral-pion
decay (\S\ref{sec:pizero}) spectra produced by homogeneous
emitting regions having nonthermal particle momentum
distributions with arbitrary shape.  Our primary goals in
developing these models were to make them as general and
accurate as possible, and to make them computationally
efficient enough to be practical for use in iterative fitting
of observational data.  An additional goal was to adhere to a
modular design to simplify customizing the models for specific
applications.  By providing alternate implementations of the
appropriate modules, one can customize the shapes of the
particle distributions and the shape of the photon spectrum
used to compute the inverse Compton model. For example,
\citet{allen:05a} used the synchrotron spectrum model with
the curved particle momentum spectrum described in
\S\ref{sec:pdf} to detect curvature in the cosmic-ray electron
spectrum of SN1006.

We assess the accuracy of the computed photon spectra by
applying a recurrence relation and by comparing with analytic
results and with published numerical models
(\S\ref{sec:accuracy}). When using these models to
fit (\S\ref{sec:observ}) simultaneously radio, x-ray and
gamma-ray observations of supernova remnants, we have
occasionally found it useful to reduce the number of degrees of
freedom by introducing additional physical constraints on the
fit parameters. Gamma-ray spectra are particularly troublesome
because they may contain significant contributions from inverse
Compton emission, neutral-pion decay and nonthermal
bremsstrahlung.  By introducing additional physical
constraints, one can usefully reduce the set of linear
combinations of these models that fit the gamma-ray data.  In
\S\ref{sec:constraints}, we discuss some of these constraints
and describe how they may be imposed.

\section{Particle Distribution Function}
\label{sec:pdf}

The algorithms used to compute the photon emission impose
relatively few limitations on distribution functions suitable
for use in fitting.  The spectral models described below are
derived assuming a particle momentum distribution function that
depends only on the magnitude of the particle momentum and
not its direction. For most practical applications, the
momentum distribution function should depend on a reasonably
small number of parameters and should be integrable by adaptive
quadrature rules.

In applications to date, we have used a nonthermal particle
distribution function of the form
\begin{equation}
N(p) = A\left(\frac{pc}{E_0}\right)^{-\Gamma + af(p)}
\exp\left(\frac{E_0-pc}{\cutoff}\right),
\label{eq:pdf}
\end{equation}
where
\begin{equation}
f(p) \equiv \left\{
\begin{array}{ll}
\log (pc/E_0), & p \ge E_0/c;\\
 0, & p < E_0/c,
\end{array}
\right.
\end{equation}
and where $p \equiv \gamma m v$, and
$E_0 \equiv 1\,\GeV$.  The same functional form is used for both
protons and electrons. The normalization parameter, $A$,
represents the density of particles with momentum $p = E_0/c$ and has
units $\cm^{-3}(\GeV/c)^{-1}$.

\begin{figure}[t]
\epsscale{1.1}
\hspace*{-1.5cm}
\plotone{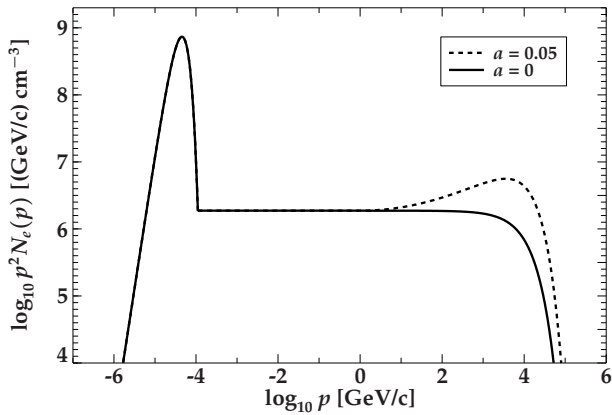}
\caption{\label{fig:density}
Sample electron momentum distribution function.  At low
energies, a Maxwellian thermal distribution ($kT = 1$\,keV) is
shown. At high energies, two nonthermal distributions are
shown, each with a high-energy exponential cutoff $\cutoff =
10$\,TeV. The dashed line shows a nonthermal distribution with
$\Gamma=2$ and with positive curvature ($a=0.05$) above
momentum $p = 1$\,GeV/c. The other nonthermal component (solid
line) has $\Gamma=2$ and zero curvature ($a=0$).
}
\end{figure}

\begin{figure}[t]
\epsscale{1.1}
\hspace*{-1.5cm}
\plotone{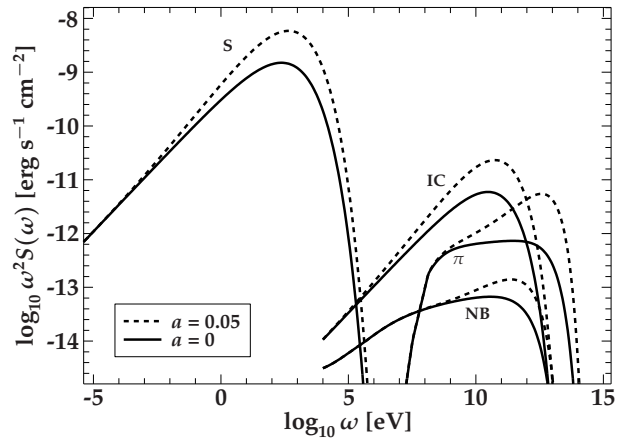}
\caption{\label{fig:curvature}
Example of the effect of curvature on nonthermal spectra.  The
solid lines show synchrotron (S), inverse Compton (IC),
neutral-pion decay ($\pi$) and nonthermal bremsstrahlung (NB)
spectra with $\Gamma=2$ and no curvature ($a = 0$).  The dashed
lines show the same spectra with positive curvature ($a=0.05$)
above particle momentum $p = 1$\,GeV/c.
\vspace*{0.5cm}
}
\end{figure}

When $a=0$, equation (\ref{eq:pdf}) describes a power-law
distribution in momentum with an exponential cutoff at $pc
\approx \cutoff$. When $a\neq 0$, the power-law exponent
changes with momentum for $p \ge E_0/c$ (see Figure
\ref{fig:density}). The effect is such that, for each factor of
ten increase in the momentum above $E_0/c$, the spectral index
changes by an amount equal to the curvature parameter, $a$.
Positive values of $a$ cause the particle spectrum slope to
flatten toward higher momenta, as shown in Figure
\ref{fig:density}. Positive values are expected due to
nonlinear behavior of the diffusive acceleration mechanism
\citep[see \eg][]{bell:87a, ellison:91a, berezhko:99a}. The
qualitative effect of a small positive curvature on the shape
of the nonthermal photon spectrum is shown in Figure
\ref{fig:curvature}.

Each of the photon emission models described below depends on
parameters associated with the underlying momentum distribution
of nonthermal particles. From equation (\ref{eq:pdf}), these parameters
are $\Gamma$, $a$ and $\cutoff$. The normalization parameter
for each photon emission model includes the normalization
parameter, $A$, for the relevant nonthermal particle
distribution function.

The low end of the particle momentum distribution is dominated
by a thermal Maxwellian as shown in Figure \ref{fig:density}.
As described in \S\ref{sec:constraints}, one can impose charge
conservation in the injection mechanism to set the value of the
cosmic-ray proton normalization, $A_p$ relative to the
cosmic-ray electron normalization, $A_e$.

In computing nonthermal photon spectra, integrals over particle
momenta generally include only relativistic particles with
$\gamma \gg 1$; in practice, we use $\gamma \ge 10$.

\vspace*{1cm}  
\section{Synchrotron Radiation}
\label{sec:sync}

From \cite{blumenthal:70a}, the total synchrotron power emitted
per unit frequency from an energetic electron ($\gamma \gg 1$)
spiraling in a magnetic field is
\begin{equation}
P_\mathrm{emitted}(\nu) =
\frac{\sqrt{3} e^3 B \sin\alpha}{m_e c^2} F\left(\frac{\nu}{\nu_c}\right),
\label{eq:p-emitted}
\end{equation}
where $e$ is the electron charge, $m_e$ is the electron mass, $B$
is the magnetic field strength, $\alpha$ is the pitch angle
between the electron's velocity vector $\vect{v}$ and the
magnetic field vector $\vect{B}$, $\nu_c$ is the critical
frequency, defined as
\begin{equation}
  \nu_c = \frac{3eB\gamma^2}{4\pi m_e c} \sin\alpha
  \equiv \nu_0 \gamma^2 \sin\alpha,
\label{eq:nu-c}
\end{equation}
and $F(x)$, the first synchrotron
function, is defined as
\begin{equation}
F(x) \equiv x \int_x^\infty \diff\xi\;K_{5/3}(\xi), \quad (x \ge 0),
\end{equation}
where $K_{5/3}(\xi)$ is an irregular modified Bessel function.
Note that equation (\ref{eq:p-emitted}) applies to frequencies well
above the gyro-frequency, where the synchrotron spectrum may be
regarded as continuous.

Assuming a steady state electron distribution, we define
$N(p, \alpha)\;\diff p\;\diff\Omega$ as the
density of nonthermal electrons with pitch angles
$\alpha$ within a solid angle $\diff\Omega$, and momenta $p$ within
$\diff p$. From \cite{blumenthal:70a},
\begin{equation}
\deriv{W}{t} = \int \diff p\int \diff\Omega\;
              P_\mathrm{emitted}(\nu)\;N(p, \alpha)
\label{eq:dwdnudt}
\end{equation}
is the total
synchrotron power received per unit volume and per unit
frequency, integrated over pitch-angles.

Combining equations (\ref{eq:p-emitted}), (\ref{eq:nu-c}) and
(\ref{eq:dwdnudt}), and assuming an isotropic distribution
of pitch-angles so that $N(p,\alpha) = N(p)/4\pi$, we obtain
\begin{multline}
\frac{\diff W}{\diff\nu \diff t} =
\frac{\sqrt{3} e^3 B}{4\pi m_e c^2}
\int \diff p\;N(p)
 \\ \times
\int_0^{2\pi} \diff\phi \int_0^\pi \diff\alpha
\sin^2\alpha \;F\left(\frac{\nu}{\nu_0\gamma^2\sin\alpha}\right),
\label{eq:sync-emis1}
\end{multline}
Expressing this result in terms of a photon emission rate per
unit energy, equation (\ref{eq:sync-emis1}) yields the differential
emissivity spectrum in the form
\begin{equation}
\frac{\diff n}{\diff \omega \diff t} =
\frac{\sqrt{3} e^3 B}{hm_e c^2\omega}
\int \diff p\;N(p)\;R\left(\frac{\omega}{\omega_0\gamma^2}\right),
\label{eq:sync-emis}
\end{equation}
where $\omega$ is the photon energy and the angular integral is
\begin{equation}
R(x) \equiv \frac{1}{2}\int_0^\pi \diff\alpha\;\sin^2\alpha\;F\left(\frac{x}{\sin\alpha}\right).
\label{eq:sync-angular-integral}
\end{equation}
\cite{crusius:86a} showed that
equation (\ref{eq:sync-angular-integral}) may be expressed analytically
in terms of Whittaker functions,
\begin{multline}
R(x) = \frac{\pi x}{2}
\left[W(x;0,4/3)W(x;0,1/3) \right. 
\\
\left. -~W(x;1/2,5/6)W(x;-1/2,5/6)\right],
\label{eq:whittaker}
\end{multline}
where $W(z;\kappa,\mu) = e^{-z/2}z^{\mu+1/2}
U\left(z;\frac{1}{2}+\mu-\kappa, 1+2\mu\right)$ and $U(z;m,n)$
is a confluent hypergeometric function of the second kind.

The synchrotron flux is obtained by evaluating equation
(\ref{eq:sync-emis}). To speed numerical computations, we
compute $R(x)$ using a cubic spline interpolation on a
pre-computed table; we use the GNU Scientific Library
\citep{galassi:05a} to perform numerical integration and
one-dimensional spline interpolation and to evaluate selected
special functions. This table is constructed by evaluating
equation (\ref{eq:whittaker}) on an adaptive grid of $x$ values
chosen to accurately sample the behavior of the function over
the range $10^{-38} < x < 100$. A log-spaced grid of $x$ values
was refined by adding interpolation points until the
interpolated value of $\log R(x)$ had a fractional error smaller
than $1.25 \times 10^{-5}$ at the midpoint of each $x$
interval; we linearly interpolated the logarithms of $x$ and
$R(x)$ only while refining the grid for the lookup table. In
the final table, the associated cubic spline interpolation
errors are $\abs{\delta R/R} \lesssim 5\times 10^{-10}$ over
the entire range of $x$.

The parameters of the synchrotron model are $\Gamma$, $a$,
$\cutoff$, the total magnetic field strength, $\btot$ and the
normalization,
\begin{equation}
 \norm_\mathrm{S} \equiv \frac{1}{4\pi d^2} \int_V \diff V A_e(r)
  = \frac{A_e V_\mathrm{S}}{4\pi d^2},
\end{equation}
where $A_e$ is the normalization of the nonthermal electron
momentum distribution, $d$ is the distance to the source and
$V_\mathrm{S}$ is the synchrotron emitting volume. A homogeneous
emitting volume is assumed.

Note that the parameters of the model describe the physical
properties of the synchrotron-emitting plasma rather than the
properties of the observed synchrotron emission. In some
situations, two or more of the fit parameters are degenerate.
For example, when fitting X-ray observations alone, the data
constrain the critical frequency, $\nu_c$, but, because
$\nu_c$ depends on the product $\btot\gamma^2$, the model
parameters $\btot$ and $\cutoff$ are degenerate.  Similarly,
when fitting radio observations alone, the normalization,
$\norm_\mathrm{S}$, and $\btot$ are degenerate.  In both cases,
freezing $\btot$ solves the problem.  Although one could group
the physical parameters to remove degeneracies for special
cases, we have chosen to keep the physical parameters separate.
The advantage of this choice is that, by introducing additional
observational constraints from other energy bands, it may
be possible to constrain the physical parameters separately.
For example, by fitting radio, X-ray and gamma-ray data
simultaneously, one might individually constrain
$\norm_\mathrm{S}$, $\Gamma$, $a$, $\cutoff$ and $\btot$.

\section{Inverse Compton Scattering}
\label{sec:invc}

Given a distribution of relativistic ($\gamma \gg 1$)
electrons $N(p)$, immersed in an isotropic radiation
field with photon number density $n(\omega_i)$, the differential
emissivity spectrum of Compton scattered photons for
single-scattering is
\begin{equation}
\frac{\diff n}{\diff \omega \diff t} = c
\int \diff \omega_i\;n(\omega_i)\;\int_{p_\mathrm{min}(\omega_i)}^\infty \diff p\;N(p)
\sigma_\mathrm{KN}(\gamma, \omega_i, \omega),
\label{eq:ic-rate}
\end{equation}
where $\omega \equiv h\nu/(m_e c^2)$, and
\begin{multline}
\sigma_\mathrm{KN}(\gamma, \omega_i, \omega) = \frac{2\pi r_0^2}{\omega_i \gamma^2}
\\ \times
\left[1 + q - 2q^2 + 2q\ln q +
\frac{\Gamma^2 q^2 (1-q)}{2(1 + \Gamma q)}
\right]
\end{multline}
is the
Klein-Nishina scattering cross-section \citep{blumenthal:70a},
where
\begin{equation}
 q \equiv \frac{\omega}{4 \omega_i \gamma (\gamma - \omega)},
\end{equation}
$\Gamma \equiv 4\omega_i\gamma$ and $r_0 = e^2 /(m_ec^2)$ is the
classical electron radius.

The allowed range for $q$ follows from the kinematics of
Compton scattering. In the frame in which the electron is
initially at rest, the incident photon has energy $\omega_i^\prime$
and the scattered photon energy may span the range $\omega_i^\prime
/ (1 + 2\omega_i^\prime) \le \omega^\prime \le \omega_i^\prime$. In the lab
frame, the scattered photon energy lies in the range $\omega_i \le \omega
\le \Gamma/(1 + \Gamma)$.  Therefore, the allowed range for $q$ is
\begin{equation}
  q_\mathrm{min} \equiv \frac{1}{4\gamma(\gamma-\omega_i)} \le q \le 1.
\end{equation}

The lower limit of the integral over electron momenta in
equation (\ref{eq:ic-rate}) corresponds to the minimum electron
momentum, $p_\mathrm{min} = \gamma_\mathrm{min}m_e v$, that can
Compton scatter a photon from initial energy $\omega_i$ to
final energy $\omega$.  From the kinematics, one can show that
the threshold electron Lorentz factor is
\begin{equation}
 \gamma_\mathrm{min} = \frac{1}{2}\left[\omega + \sqrt{\omega^2 + \frac{\omega}{\omega_i}}\right],
\label{eq:ic-threshold}
\end{equation}
which corresponds to $q=1$.

The inverse Compton model is obtained by evaluating
equation (\ref{eq:ic-rate}) for the cosmic background radiation
field so that
\begin{equation}
  n(\omega_i) = \frac{1}{\pi^2 \lambda^3}
  \frac{\omega_i^2}{e^{\omega_i/\Theta} - 1},
\end{equation}
where $\lambda \equiv \hbar/(m_e c)$ is the electron Compton
wavelength, $\Theta \equiv kT/(m_e c^2)$, and $T$ = 2.725\,K
\citep{bennett:03a}.

For computational convenience, it is useful to change the order
of integration.  Performing the integral over photon energies
first, we let the electron momentum integral extend over
the full range of electron momenta and use the cross-section
\begin{equation}
\sigma(\gamma, \omega_i, \omega) = \left\{
\begin{array}{ll}
\sigma_\mathrm{KN}(\gamma, \omega_i, \omega), & q_\mathrm{min} \le q \le 1; \\
0,& {\rm otherwise}.
\end{array}
\right.
\end{equation}
We can then speed up numerical
computations by tabulating the photon energy integral,
\begin{equation}
I_c(\omega, \gamma) \equiv \int
\diff\omega_i\;n(\omega_i)\;\sigma_\mathrm{KN}(\gamma, \omega_i, \omega),
\label{radiation-field-integral}
\end{equation}
for a given spectrum of seed photons, $n(\omega_i)$.  In
practice, we use $q$ as the variable of integration to evaluate
the integral in equation (\ref{radiation-field-integral}).

Because the range of $I_c(\omega, \gamma)$ spans many orders of
magnitude over the domain of interest, we find it useful to
interpolate on the logarithm of the function value. In
constructing an interpolation table for $I_c(\omega,\gamma)$,
we explicitly incorporate the existence of the threshold
Lorentz factor $\gamma_\mathrm{min}$.  Because $I_c(\omega,
\gamma)$ asymptotically goes to zero as $\gamma \rightarrow
\gamma_\mathrm{min}$, we set $I_c(\omega,\gamma) = 0$ for
$\gamma < \gamma_0 \equiv
\gamma_\mathrm{min}(\omega_{i,\mathrm{max}})$ where
$\omega_{i,\mathrm{max}}$ is the maximum incident photon energy
of interest. Introducing a change of variables, $x_c \equiv
\log(\log (\gamma/\gamma_0))$ and $y_c \equiv \log \omega$, we
tabulate $\log I_c(x_c, y_c)$ on a $1024\times 1024$ uniform
rectangular grid in $x_c$ and $y_c$ covering the range $10 \le
\gamma \le 10^{9.5}$ and $10^2 \le \omega \le 10^{15}$\,eV. The
smoothness of the resulting table allows accurate
two-dimensional interpolation with a $6^\mathrm{th}$-order
spline \citep{deboor:78a} in each coordinate.

The parameters of the inverse Compton model are $\Gamma$, $a$,
$\cutoff$, the blackbody temperature, $T$ and
the normalization,
\begin{equation}
 \norm_\mathrm{I} \equiv \frac{1}{4\pi d^2} \int_V \diff V A_e(r)
  = \frac{A_e V_\mathrm{I}}{4\pi d^2},
\end{equation}
where $V_\mathrm{I}$ is the homogeneous emitting volume. By default, the
temperature parameter, $T$, is not used and the model uses a
lookup table for $I_c(\omega,\gamma)$ appropriate for seed
photons from the cosmic background radiation field. The
error associated with interpolation in this table is typically
$\abs{\delta I_c/I_c} < 10^{-10}$.

A switch is provided to force the model to compute the
radiation field integral by direct integration instead of table
interpolation. The computational expense of direct integration
currently makes this mode impractical for use in spectral
fitting.

Alternatively, one can use a linear combination of lookup
tables to describe a more complicated radiation field. For
example, one can construct lookup tables corresponding to a
dilute stellar radiation field or to emission from molecular
clouds. During the fit, one can vary the relative proportions
of these components.  Note that the current implementation
restricts the input radiation field shape to one that is
integrable by adaptive quadrature rules.

The inverse Compton emitting volume, $V_\mathrm{I}$, need not
be the same as the synchrotron emitting volume, $V_\mathrm{S}$.
Because the cosmic background radiation photons will fill the
entire synchrotron emitting volume, $V_\mathrm{I}$ will usually
be at least as large as $V_\mathrm{S}$. However, if nonthermal
electrons are found in a volume with a relatively weak magnetic
field, that volume will produce inverse Compton emission, but
little synchrotron emission, and $V_\mathrm{I} > V_\mathrm{S}$.
For this reason, it is useful to allow the inverse Compton and
synchrotron norms to be different.

\section{Nonthermal Bremsstrahlung}
\label{sec:ntbrem}

In this section, we consider nonthermal bremsstrahlung emission
from a population of nonthermal electrons incident on a
stationary target containing free electrons and ions.  The
total bremsstrahlung emissivity is computed as a sum of
contributions from \ee\ and \ez\ bremsstrahlung.

Given two populations of relativistic particles with momentum
distributions, $N_1(p_1)$ and $N_2(p_2)$, and with interaction
cross-section, $\sigma$, the general expression for the collision
rate per unit volume is
\begin{multline}
 \deriv{n}{t} = \left(1+\delta_{12}\right)^{-1}
 \int \diff p_1 N_1(p_1) \int \diff p_2 N_2(p_2)
 \\ \times
 \;\sigma \sqrt{\left(\vect{v_1} - \vect{v_2}\right)^2 -
       \left(\vect{v_1}\times\vect{v_2}\right)^2/c^2},
\label{eq:general-rate}
\end{multline}
where the integrals extend over the momenta of the interacting
distributions \citep{landau:75a}.  The Kronecker delta,
$\delta_{12}$ corrects for double-counting when the particles
are identical.

When $\abs{\vect{v_2}} \ll \abs{\vect{v_1}}$, equation
(\ref{eq:general-rate}) reduces to the familiar non-relativistic
form, and one of the populations may be treated as a stationary
target. For example, when the target particles may be
characterized by a Maxwellian thermal distribution, the thermal
motions of the target particles may be neglected as long as
$kT_2 \ll (\gamma-1) mc^2$.

\subsection{Electron-Electron Bremsstrahlung}

In the limit that $\abs{\vect{v_2}} \ll \abs{\vect{v_1}}$,
equation (\ref{eq:general-rate}) yields a differential emissivity
spectrum for \ee\ bremsstrahlung of the form
\begin{equation}
\frac{\diff n}{\diff \omega \diff t} = n_e \int
\diff{p}\;N(p)\;v \deriv{\sigma_{ee}}{\omega},
\label{eebrems-spectrum}
\end{equation}
where $n_e$ is the target electron density, and
$\diff\sigma_{ee}/\diff \omega$ is
the lab-frame differential cross-section for this interaction
between identical particles and includes the
factor $\left(1+\delta_{12}\right)^{-1}$.

The lab-frame cross-section,
$\diff\sigma_{ee}/\diff\omega\diff\psi$, differential in both
photon energy, $\omega$, and photon emission angle, $\psi$ was
taken from \cite{haug:75a} and was computed using software
kindly provided by E. Haug. The lab-frame cross-section,
differential in photon energy, $\diff\sigma_{ee}/\diff\omega$,
is obtained by integrating over photon emission angles. The
angular integration limits are given by \cite{haug:75a} and
follow from energy-momentum conservation. In the center of
momentum (CM) frame, the photon emission is symmetric along the
direction of motion and all values of $\psi_*$ are accessible.
For an electron incident with momentum $p = \gamma m v$ in the
lab frame, beaming restricts photon emission angles to a narrow
cone in the forward direction. From \cite{haug:75a}, photons
with energies $(\gamma -1)/(\gamma + \gamma \beta + 1) < \omega
\le (\gamma -1)/(\gamma - \gamma \beta + 1)$ are emitted into a
cone with the maximum emission angle, $\psi_\mathrm{max}$,
given by
\begin{equation}
\cos {\psi_\mathrm{max}} = \frac{(\gamma + 1)\omega - (\gamma-1)}{\omega
\gamma \beta}.
\end{equation}
Lower energy photons may span $0 \le \psi \le \pi$.

In carrying out the angular integration for ultra-relativistic
electrons, some care must be taken to minimize numerical
problems due to round-off errors. In particular, the expression
for the differential cross-section, $\diff\sigma_{ee}/\diff
\omega \diff \psi$, includes terms that divide by the quantity
\begin{equation}
  x = \omega (E - p\cos\psi),
\end{equation}
where $E \equiv T + 1$, $p \equiv \sqrt{T(T+2)}$, and $T$ is
the incident electron kinetic energy in units of $m_e c^2$. For
$\gamma > 10^8$, $x$ is identically zero in double-precision
for $\cos\psi=1$, causing a division by zero error.  In the
ultra-relativistic regime, we computed $x$ using the first few
terms of its series expansion in powers of $\gamma^{-2}$.
Substituting $\cos\psi \equiv 1 - s$, and using $s$ as the
variable of integration, the relevant cancellations can be
handled analytically. To check the result, we used the
Lorentz invariant, $\omega \diff^3\sigma/\diff p^3$, to
transform Haug's CM-frame cross-section into the lab frame
\citep[see \eg][]{dermer:86a}.  In the CM frame, the cross-section
computation is less affected by round-off errors because the
relevant electron Lorentz factor is $\gamma_c = ((\gamma +
1)/2)^{1/2}$. Applying the Lorentz transformation and treating
round-off errors by handling the relevant cancellations
analytically, we verified that the transformed result agreed
with the lab-frame cross-section in the ultra-relativistic
regime.

To speed up the numerical integrations, the cross-section is
evaluated using a two-dimensional cubic-spline interpolation on
pre-computed tables. In constructing an interpolation table for
$\diff\sigma_{ee}/\diff \omega \equiv \sigma^\prime$, we
explicitly incorporate the fact that $\sigma^\prime(T,\omega)
\equiv 0$ for kinetic energies $T < T_\mathrm{min}(\omega)$,
where $T_\mathrm{min}(\omega)$ is determined by the kinematics.
We define $T_\mathrm{min}(\omega)$ numerically as the locus of
points at which $\sigma^\prime(T_\mathrm{min},\omega) =
\epsilon_B \mathrm{max}\{\sigma^\prime(T,\omega)\}$ for
$\epsilon_B = 10^{-8}$. Introducing a change of variables, $x_B
\equiv \log(\log (T/T_\mathrm{min}))$ and $y_B \equiv \log
\omega$, we tabulate $\log \sigma^\prime(x_B, y_B)$ on a
$1024\times 1024$ rectangular grid in $x_B$ and $y_B$ covering
$10^2 \le T/mc^2 \le 10^{9.5}$ and $10^2 \le \omega/mc^2 \le 10^{9.5}$.
The smoothness of the resulting table allows accurate two-dimensional
cubic spline interpolation. The error associated with interpolation in the
cross-section table is $\abs{\delta
\sigma^\prime/\sigma^\prime} < 10^{-5}$.

\subsection{Electron-Ion Bremsstrahlung}

From equation (\ref{eq:general-rate}) the differential emissivity
spectrum for \ez\ bremsstrahlung can be written
\begin{equation}
\frac{\diff n}{\diff \omega \diff t} = n_Z \int
\diff{p}\;N(p)\;v\deriv{\sigma_{eZ}}{\omega},
\label{eq:eZ-brems}
\end{equation}
where $n_Z$ is the density of target ions with charge
$Z$, and $\diff\sigma_{eZ}/\diff \omega$ is the lab-frame differential
cross-section.

The Bethe-Heitler cross-section \citep{heitler:53a, koch:59a}
determines the probability that deflection of a relativistic
electron in the unscreened field of an ion of charge $Z$ will
yield a photon of energy $\omega = h\nu/(m_ec^2)$.  The
differential cross-section may be written in the form
\begin{multline}
\deriv{\sigma_{eZ}}{\omega} =
\frac{\bar{\phi}}{\omega}
\frac{\gamma\beta}{\gamma_0\beta_0}
\left[
\frac{4}{3}
- \frac{2\gamma_0\gamma(\gamma_0^2\beta_0^2
+ \gamma^2\beta^2)}{\gamma_0^2\gamma^2\beta_0^2\beta^2} \right.
\\
\left.
+\;\frac{a_0\gamma  }{\gamma_0^3\beta_0^3}
+ \frac{a\gamma_0}{\gamma^3\beta^3}
- \frac{a_0a}{\gamma_0\gamma\beta_0\beta}
+ L x
\right],
\label{eq:heitler-xsec}
\end{multline}
where
\begin{multline}
x \equiv \frac{8}{3\beta_0\beta} +
\frac{\omega^2(1+\beta_0^2\beta^2)}{\gamma_0\gamma\beta_0^3\beta^3}
   + \frac{\omega}{2\gamma_0\gamma\beta_0\beta}
   \\ \times
   \left[
         \frac{a_0(\gamma   +
\gamma_0\beta_0^2)}{\gamma_0^2\beta_0^3}
          - \frac{a  (\gamma_0 + \gamma  \beta^2  )}{\gamma^2  \beta^3  }
          + \frac{2\omega}{\gamma_0\gamma\beta_0^2\beta^2}
                                         \right],
\end{multline}
\begin{equation}
\bar{\phi} \equiv \alpha Z^2 r_0^2, \quad r_0 = e^2/(m_ec^2),
\end{equation}
\begin{equation}
a_0 \equiv 2\ln\left[\gamma_0\left(1+\beta_0\right)\right], \quad
a \equiv 2\ln\left[\gamma  \left(1+\beta  \right)\right],
\end{equation}
\begin{equation}
 L \equiv 2\ln\left[\left(\gamma_0\gamma +
\gamma_0\gamma\beta_0\beta - 1\right)/\omega\right].
\end{equation}
In these expressions, $\alpha$ is the fine structure constant
and $\gamma(\beta)$ and $\gamma_0(\beta_0)$ are the Lorentz
factors of the scattered and incident electron, respectively.
The Bethe-Heitler cross-section is derived using the Born
approximation which is appropriate in the limit of high kinetic
energies, such that $2\pi\alpha Z/\beta \ll 1$ for both the
incident and scattered electron. The recoil of the nucleus is
neglected so $\gamma_0(\beta_0) = \gamma(\beta) + \omega$.  The
accuracy at low energies is improved by including the Elwert
correction factor, which we apply at all energies
\citep{elwert:39a,pratt:75a,haug:97a}, so that
$\diff\sigma_{eZ}/\diff\omega \rightarrow
\eta_E\diff\sigma_{eZ}/\diff\omega$ where
\begin{equation}
\eta_E \equiv \frac{\xi}{\xi_0}\frac{1 - \exp(-\xi_0)}{1 - \exp(-\xi)},
\end{equation}
and where
\begin{equation}
\xi \equiv 2\pi\alpha \frac{Z}{\beta}, \quad \xi_0 \equiv
2\pi\alpha \frac{Z}{\beta_0}.
\end{equation}
When the target ions retain bound atomic electrons, it is
necessary to modify the cross-section to account for screening
of the nuclear charge.  We assume that the target material is
completely ionized so that screening corrections are not
necessary.

The relative simplicity of the cross-section makes it practical
to perform the integration over electron momenta by evaluating
(\ref{eq:heitler-xsec}) directly rather than interpolating
values from a pre-computed table.

The parameters of the nonthermal bremsstrahlung model are $\Gamma$, $a$,
$\cutoff$ and the normalization,
\begin{equation}
 \norm_\mathrm{B} \equiv \frac{1}{4\pi d^2} \int_V \diff V\;n_0(r)A_e(r)
  = \frac{A_e n_0 V_\mathrm{B}}{4\pi d^2},
\end{equation}
where $n_0 \equiv \sum_Z n_Z$ is the total ion number density
of the target, $n_Z$ is the number density of ions with charge
$Z$ and $V_\mathrm{B}$ is the emitting volume.  The user
interface includes parameters to control the relative
contributions of \ee\ and electron-proton bremsstrahlung.  The
\ee\ contribution has weight $X_e \equiv \sum_Z Z n_Z/n_0$ and
the electron-proton contribution has weight $X_i \equiv \sum_Z
Z^2 n_Z/n_0$. By default, the target is assumed to consist of
hydrogen and helium with $n_\mathrm{He}/n_\mathrm{H}=0.1$ so
that the default weights are $X_e = X_\mathrm{H} +
2X_\mathrm{He} = 1.091$ and $X_i = X_\mathrm{H} +
4X_\mathrm{He} = 1.273$. A common alternative is to view $n_0$
as the target proton density and to compute the weights using
that assumption.  As long as the weights are consistent with
the definition of $n_0$, the result is the same.

\section{Neutral Pion Decay}
\label{sec:pizero}

Nonthermal protons produce gamma-ray emission primarily through
collisions with thermal protons.  These collisions yield
neutral pions via $pp \rightarrow \pi^0 + X$, and the neutral
pions decay via $\pi^0 \rightarrow 2\gamma$. We include only
the contribution from proton-proton collisions and ignore
contributions from processes other than the decay of neutral
pions.  In its rest-frame, the neutral pion decays within $\sim
10^{-16}$ sec, producing two gamma-rays, each with an energy of
$\omega_0 = \frac{1}{2}m_\pi c^2$.  Because the pion has zero
spin, the gamma-rays are emitted isotropically.  Following
\cite{hillier:84a}, the number of gamma-rays emitted into a
rest frame angle between $\theta_*$ and $\theta_* +
\diff\theta_*$ is
\begin{equation}
n(\theta_*)\diff\theta_* = \sin \theta_*\diff\theta_*.
\end{equation}
Since a gamma-ray that is emitted at an angle
$\theta_*$ in the pion rest frame has a lab-frame energy
\begin{equation}
 \omega = \omega_0 \gamma_\pi \left(1 - \beta_\pi \cos
 \theta_*\right),
\label{eq:labframe-gamma-energy}
\end{equation}
the number of gamma-rays with lab-frame
energy between $\omega$ and $\omega + \diff\omega$ is
\begin{equation}
n(\omega) = n(\theta_*)\deriv{\theta_*}{\omega} =
\frac{1}{\omega_0\gamma_\pi\beta_\pi} = \frac{2}{p_\pi c},
\end{equation}
where $p_\pi = \gamma_\pi m_\pi v_\pi$ is the pion momentum.
For a population of pions with momentum distribution,
$N_\pi(p_\pi)$, the gamma ray spectrum
is then
\begin{equation}
  n(\omega) = 2\int_{p_{\pi,\mathrm{min}}}^{p_{\pi,\mathrm{max}}}
  N_\pi(p_\pi)\;\frac{\diff p_\pi}{p_\pi}.
\end{equation}
The lower integration limit is set by the minimum pion momentum
required to yield a lab-frame photon of energy $\omega$.  From
equation (\ref{eq:labframe-gamma-energy}), the low energy
gamma-ray has energy $\omega_\mathrm{min} = \omega_0 \gamma_\pi
(1 - \beta_\pi)$, corresponding to a pion velocity of
\begin{equation}
\beta_{\pi, \mathrm{min}} = \frac{\omega_0^2 - \omega^2}{\omega_0^2 + \omega^2}.
\end{equation}
A lab frame photon of energy
$\omega$ therefore requires a pion Lorentz factor of at least
\begin{equation}
 \gamma_{\pi,\mathrm{min}} =
 \frac{1}{2}\left(\frac{\omega}{\omega_0}
 + \frac{\omega_0}{\omega}\right).
\end{equation}
The upper integration limit corresponds to the maximum pion
momentum that can be produced by a proton of kinetic energy
$T_{p, \mathrm{max}}$.  From the collision kinematics,
one can show that
\begin{equation}
\gamma_{\pi,\mathrm{max}} =
\frac{1}{2}\left(\frac{T_{p,\mathrm{max}}}{2\omega_0}
+ \frac{\omega_0}{m_p c^2}\right).
\end{equation}

The production spectrum of secondary pions in proton-proton
collisions is
\begin{equation}
N_\pi(p_\pi) = n_p \int_{p_{\mathrm{min}}(p_\pi)}^\infty
 \diff p\;v_p N_p(p) \deriv{\sigma(p_\pi, p)}{p_\pi},
\end{equation}
where $n_p$ is the density of target protons, $v_p$ is the
nonthermal proton velocity, $N_p(p)$ is the nonthermal proton
momentum distribution, and $\diff\sigma(p_\pi, p)/\diff p_\pi$
is the differential cross-section for production of a neutral
pion with lab-frame momentum $p_\pi$ from a proton with
lab-frame momentum $p$. The lower integration limit is the
threshold proton momentum for producing a pion with momentum
$p_\pi$.  From the collision kinematics, one can show that the
threshold proton kinetic energy is
\begin{equation}
  T_{p,\mathrm{min}}(p_\pi) =
  2\left(p_\pi^2c^2 + m_\pi^2c^4\right)^{1/2}
  + \frac{m_\pi^2 c^2}{2 m_p}.
\end{equation}

For the purpose of interactive spectral fitting, computing the
photon spectrum from neutral-pion decay by direct integration
over the proton momentum distribution and over the resulting
pion spectrum is quite computationally demanding. The cost of
the numerical integrations is compounded by the fact that the
differential cross-sections are computationally expensive for
certain energies \citep[see \eg][]{dermer:86a, mori:97a}. To
reduce the cost of these computations, we evaluate the integral
over the proton distribution using the delta-function
approximation described by \cite{aharonian:00a}.

The parameters of the neutral-pion decay model are $\Gamma$, $a$,
$\cutoff$ and the normalization,
\begin{equation}
 \norm_\pi \equiv \frac{1}{4\pi d^2} \int_V \diff V\;n_p(r)A_p(r)
  = \frac{A_p n_p V_\pi}{4\pi d^2},
\end{equation}
where $A_p$ is the normalization for the nonthermal proton
momentum distribution, $n_p$ is the density of target protons,
and $V_\pi$ is the emitting volume.  Note that for this model,
the parameters $\Gamma$, $a$ and $\cutoff$ refer to the proton
momentum distribution.

\begin{figure}[t]
\epsscale{1.1}
\hspace*{-1.5cm}
\plotone{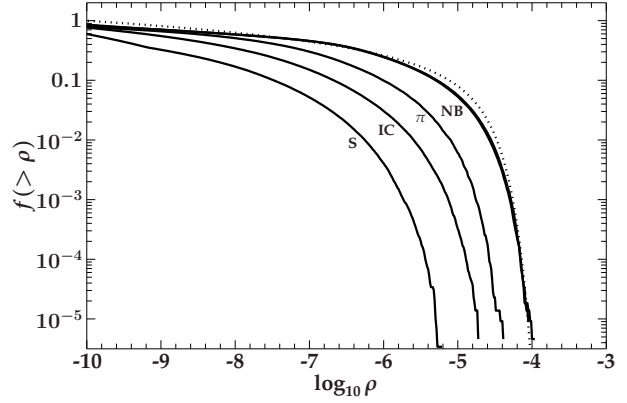}
\caption{\label{fig:recur-distrib}
Distribution of fractional recurrence relation errors.  The
fraction, $f$, of errors larger than $\recur$, is shown for
each spectral model. The errors were computed at many points
along each of several model spectra.  One spectrum was
generated for each pair of values, $\Gamma$ and $\cutoff$, over
a grid spanning $1.8 \le \Gamma \le 4$ and $1 \le \cutoff \le
1000$\,TeV. The dotted line corresponds to
$f(>\recur) = \exp(-\recur^2/2\sigma^2)$, with $\sigma=2.5
\times 10^{-5}$.  Note that the curves for \ee\ and \ep\
bremsstrahlung are almost identical.
}
\end{figure}

\section{Accuracy of Computed Spectra}
\label{sec:accuracy}

Consider spectral models of the form
\begin{equation}
 S_\Gamma(\omega; \varepsilon_c) = \int \diff p
\left(\frac{p}{\varepsilon_0}\right)^{-\Gamma}
\exp\left(\frac{\varepsilon_0-p}{\varepsilon_c}\right)\;\phi(p,\omega),
\label{eq:recur-direct}
\end{equation}
where $\omega$ is the photon energy, $p$ is the particle
momentum, $\varepsilon_0$
is a scaling constant, and $\phi(p,\omega)$ depends on the
physical process. Taking the derivative with respect to
$1/\varepsilon_c \equiv \alpha$, we find that
\begin{equation}
\frac{1}{\varepsilon_0}\deriv{S_\Gamma(\omega; \alpha)}{\alpha} =
 S_\Gamma(\omega; \alpha) -
 S_{\Gamma - 1}(\omega; \alpha).
\label{eq:recur}
\end{equation}
This identity specifies a recurrence relation between spectra
of the form shown in equation (\ref{eq:recur-direct}).

Numerical evaluation of equation (\ref{eq:recur-direct}) yields model
spectra of the form $S_\Gamma(\omega; \alpha) + \Lambda(\omega;
\alpha, \Gamma)$, where $S_\Gamma$ is the exact result and
$\Lambda$ represents the error in the computation. Because
accurately computed spectra must satisfy equation (\ref{eq:recur}), it
follows that $\Lambda(\omega; \alpha, \Gamma)$ must also
satisfy equation (\ref{eq:recur}).  Although an error in the computation
of $\phi(p,\omega)$ might generate such $(\alpha,\Gamma)$ dependent
errors in the computed spectrum, such an error should be
detectable on comparison with an analytic solution or with an
independent numerical calculation.  Eliminating that
possibility, the error term in computed spectra that satisfy
equation (\ref{eq:recur}) must be independent of $\alpha$ and
$\Gamma$ so that $\Lambda(\omega; \alpha, \Gamma) =
\Lambda(\omega)$.  The magnitude of the energy-dependent error,
$\Lambda(\omega)$ must be estimated by other means, such as by
comparing with analytic solutions and with independent
numerical results.  Although satisfying the recurrence relation
does not prove that the spectral computations are correct, it
does provide strong constraints on the magnitude and parameter
dependence of the error term, $\Lambda$.

To verify equation (\ref{eq:recur}) numerically, we introduce a
finite-difference approximation for the derivative. It follows
that computed spectra of the form shown in equation
(\ref{eq:recur-direct}) should obey a relationship of the form
$\Recur(\omega; \Gamma, \alpha) = 0$ where
\begin{multline}
\Recur(\omega; \Gamma, \alpha) \equiv
1 - \frac{S_{\Gamma-1}(\alpha)}{S_\Gamma(\alpha)} 
\\
- \lim_{\delta\alpha \rightarrow 0}
\frac{S_\Gamma(\alpha + \delta\alpha/2) - S_\Gamma(\alpha-\delta\alpha/2)}{
S_\Gamma(\alpha) \varepsilon_0 \delta\alpha}.
\label{eq:recur-finite}
\end{multline}
For clarity in equation (\ref{eq:recur-finite}), we have avoided writing
out the explicit dependence of the right-hand side upon photon
energy, $\omega$.  The centered difference used to approximate
the derivative in equation (\ref{eq:recur-finite}) should yield quadratic
convergence in the limit that $\delta\alpha \rightarrow 0$. For
each of our spectral models, we verified that, in the limit
$\delta\alpha\rightarrow 0$, $\Recur \rightarrow 0$ as $\Recur
\propto (\delta\alpha)^2$; observing smooth convergence at the
expected rate confirms that the models behave as expected. In
subsequent evaluations of equation (\ref{eq:recur-finite}), we adopt a value of
$\delta\alpha/\alpha = 2.5\times10^{-6}$. Because individual
terms in equation (\ref{eq:recur-finite}) can be $\gg 1$, it is useful
to compute the fractional error, $\recur \equiv \Recur /
\vert\Recur\vert$, where the denominator is the $\ell^2$-norm
of the three terms in equation (\ref{eq:recur-finite}).

To test the $\Gamma$ and $\alpha$ dependence of $\Lambda$, we
computed values of $\recur$ at many points along each of
several spectra.  One spectrum was computed for each pair of
($\Gamma$, $\alpha$) values on a grid sampling the range
$1.8\le \Gamma \le 4$ and $1 \le \cutoff \le 1000$\,TeV.  For
the synchrotron model, $\recur$ was evaluated at photon
energies in the range $10^{-5}\,{\rm eV} \le \omega \le
10^5\omega_c$, where $\omega_c$ is the critical energy. For the
other models, $\recur$ was evaluated at photon energies in the
range $10^5\,{\rm eV} \le \omega \le 3\cutoff$. Each photon
energy grid was logarithmically spaced and used 80 points per
decade.  Figure \ref{fig:recur-distrib} shows that, throughout
this range, $\recur$ was $\lesssim 10^{-4}$. We conclude
that, throughout the parameter range of interest, the $(\alpha,
\Gamma)$ dependence of $\Lambda$ is no larger than about a part
in $10^4$.  This result suggests that the errors are
essentially independent of the parameters of the particle
distribution function.

To test the dependence of the errors on the other parameters,
including the photon energy, we compare our results with other
analytic and numerical solutions. \cite{blumenthal:70a} derive
the well-known analytic result that the synchrotron spectrum
from a power-law distribution of electrons, $N(\gamma) \propto
\gamma^{-\Gamma}$, is itself a power-law of the form $S(\nu)
\propto \nu^{-(\Gamma-1)/2}$. To compare with this result, we
used a power-law electron distribution function to compute
numerically $S_\mathrm{sync}(\nu) \equiv \omega\diff n/\diff
\omega\diff t$, as described above in equation
(\ref{eq:sync-emis}), and we computed $S(\nu) \equiv \diff
W/\diff\nu\diff t$ from equation (4.59) of
\cite{blumenthal:70a}.  In deriving the analytic solution, the
lower limit of the integral over $\gamma$ is extended to zero
\citep{blumenthal:70a}, effectively neglecting the electron
mass. Consistent with this assumption, the value of
$x=\nu/\nu_c$ in the numerical integrand must be computed using
the approximation $\gamma \approx p/(m_e c)$ otherwise, the
computed spectrum departs from the analytic solution at low
frequencies. We find that the absolute value of the fractional
error is
\begin{equation}
  \abs{\epsilon_\mathrm{sync}} \equiv \abs{1 - \frac{S_\mathrm{sync}(\nu)}{S(\nu)}}
  < 3 \times 10^{-11},
\end{equation}
for frequencies in the range $10^7 \le \nu \le 10^{20}$\,Hz and
for values of $\Gamma$ and $\btot$ in the ranges $1 \le \Gamma
\le 4$ and $1 \le \btot \le 10^4\,\mu$G, respectively. The
primary source of error in this comparison is associated with
interpolation of $R(x)$ in our precomputed table; computing
$R(x)$ directly in terms of hypergeometric functions, we
reproduce the analytic solution to within
$\abs{\epsilon_\mathrm{sync}}
< 3 \times 10^{-13}$ over the specified range of $\nu$,
$\Gamma$ and $\btot$.

\begin{figure}[t]
\epsscale{1.1}
\hspace*{-1.5cm}
\plotone{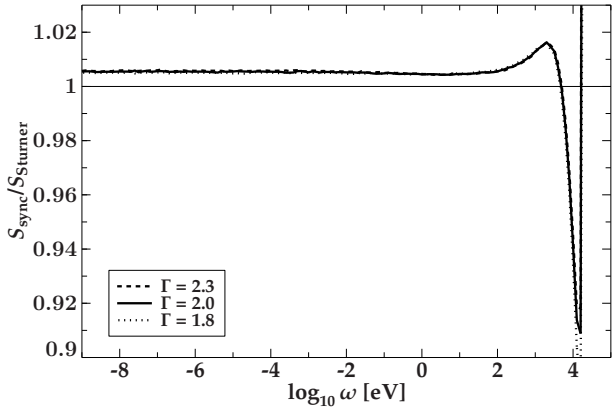}
\caption{\label{fig:sturner-sync-diff}
Ratio showing our synchrotron models divided by those of
\citet{sturner:97a} for $\Gamma=1.8$
(dashed), $\Gamma=2$ (solid), and $\Gamma=2.3$ (dotted).  Note
that the three curves are almost identical.
}
\end{figure}

\begin{figure}[t]
\epsscale{1.1}
\hspace*{-1.5cm}
\plotone{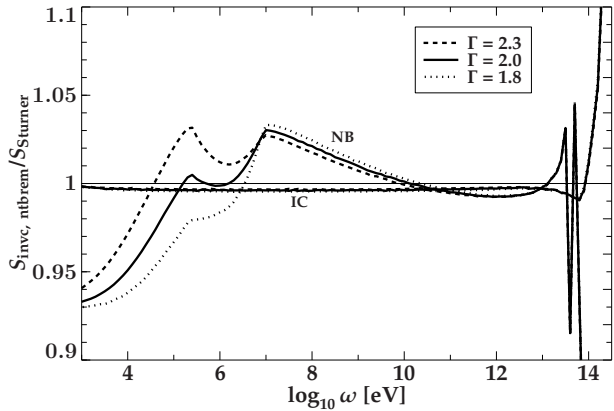}
\caption{\label{fig:sturner-gamma-diff}
Ratio showing our inverse Compton and
nonthermal bremsstrahlung spectra divided by those of
\citet{sturner:97a} for $\Gamma=1.8$ (dashed),
$\Gamma=2$ (solid), and $\Gamma=2.3$ (dotted).  The three
curves for the inverse Compton spectra are almost identical.
}
\end{figure}

\cite{blumenthal:70a} also discuss the inverse Compton spectrum
produced by a power-law distribution of electrons scattering
photons from a blackbody radiation field.  They give asymptotic
analytic solutions valid in the Thomson limit and in the
extreme Klein-Nishina limit. To compare our inverse Compton
model with these analytic solutions, we used a power-law
electron distribution function, $N(\gamma) \propto
\gamma^{-\Gamma}$, to compute numerically
$s_\mathrm{invc}(\omega) \equiv \diff n/\diff \omega\diff t$,
as described above in equation (\ref{eq:ic-rate}).  For the
purpose of these tests, we evaluated the integral over incident
photon energies (equation \ref{radiation-field-integral}) by
direct numerical integration rather than by spline
interpolation in precomputed tables.  

In the Thomson limit, the energy of the incident photon in the
electron rest frame is small compared to the electron rest
energy and the momentum transferred by the electron is small
compared to its initial momentum.  In this regime,
corresponding to $\omega_i \ll \omega \ll \gamma mc^2$, the
scattering cross-section is essentially independent of the
energy of the incoming photon. The Thomson limit approximation
is most accurate in the limit of low radiation temperature,
$T$, and for scattered photon energies near the middle of the
applicable energy range. Such a case provides the best test of
the accuracy of our inverse Compton model in the Thomson limit.
For power-law slopes in the range $1 \le \Gamma \le 4$ and
radiation field temperatures $10^{-2} \le T \le 10^4$\,K, we
used our model to compute inverse Compton spectra,
$s_\mathrm{invc}(\omega)$, for scattered photon energies in the
range $10\gamma_\mathrm{min}^2 kT \le \omega \le 0.1
\gamma_\mathrm{min} mc^2$, using $\gamma_\mathrm{min} \equiv
10$.  For comparison, we computed $s(\omega) \equiv \diff
N_\mathrm{tot}/\diff t\diff \epsilon_1 \propto
\omega^{-(\Gamma+1)/2}$ using equation (2.65) of
\cite{blumenthal:70a}. We found that our inverse Compton model
converged smoothly to the analytic result in the appropriate
limit.  The smallest fractional difference is
\begin{equation}
\abs{\epsilon_\mathrm{invc}} \equiv 
\abs{1 - \frac{s_\mathrm{invc}(\omega)}{s(\omega)}} < 10^{-8}.
\end{equation}
Broader comparisons are less useful as a test of computational
accuracy because the asymptotic solution itself becomes less
accurate with increasing $T$ and for energies approaching the
endpoints of the applicable range.  For $T=0.01$\,K, the
fractional error is smallest ($\abs{\epsilon_\mathrm{invc}} <
10^{-8}$) at $\omega \approx 50$\,eV, then increases toward
lower and higher energies with $\abs{\epsilon_\mathrm{invc}} <
10^{-3}$ for the range $10^{-2} \le \omega \le 10^7$\,eV.
Increasing $T$ narrows the applicable energy range so that for
$T=10^4$\,K, $\abs{\epsilon_\mathrm{invc}} < 10^{-3}$ only for
the range $2 \lesssim \omega \lesssim 5$\,keV.

In the extreme Klein-Nishina limit, the energy of the incident
photon in the electron rest frame is large compared to the
electron rest energy.  In this regime, corresponding to
$\omega_i \gamma \gg mc^2$, the scattering cross-section is
strongly peaked near the maximum scattered photon energy so
that individual Compton scatterings tend to involve a large
energy transfer.  The characteristic scattered photon energy is
then $\omega \sim \gamma mc^2$.  It follows that this regime
may be characterized by the requirement that $\omega_i \omega
\gg m^2c^4$ so that, for a blackbody radiation field with
$\omega_i \sim kT$, the extreme Klein-Nishina limit corresponds
to scattered photon energies $\omega \gg m^2c^4/(kT)$. We
computed $s(\omega) \equiv \diff N_\mathrm{tot}/\diff t\diff
\epsilon_1$ using equation (2.88) of
\cite{blumenthal:70a}\footnote{Equation (2.88) of
\cite{blumenthal:70a} involves a constant, $C_l$, that is
defined in terms of an infinite series that converges extremely
slowly: $C_l \equiv \left(6/\pi^2\right)\sum_{k=1}^\infty \ln
k/k^2 \approx 0.5700$. From its series definition, it follows
that this constant is equivalent to $C_l =
-\left(6/\pi^2\right)\zeta^\prime(2)$, where $\zeta^\prime$ is
the first derivative of the Riemann zeta function.  To enable
more precise quantative comparison with the analytic solution,
we used the numerical value $C_l = 0.56996099309453280$,
obtained using the symbolic algebra package {\sc maple}.}. In
the extreme Klein-Nishina limit, we computed inverse Compton
spectra, $s_\mathrm{invc}(\omega)$, for scattered photon
energies in the range $10^3 m^2c^4 /(kT) \le \omega \le 10^{12}
m^2c^4 /(kT)$ for power-law slopes in the range $1 \le \Gamma
\le 4$ and radiation field temperatures in the range $T=10^3 -
10^8$\,K. We found that the inverse Compton model spectra
converged smoothly to the analytic result, $s(\omega)$. The
fractional error decreased smoothly with increasing photon
energy from $\abs{\epsilon_\mathrm{invc}} \approx 10^{-3}$ at
$\omega \approx 10^3 m^2c^4 /(kT)$ to
$\abs{\epsilon_\mathrm{invc}} \approx 3 \times 10^{-8}$ for
$\omega \approx 10^9 m^2c^4 /(kT)$. Numerical round-off errors
become important at higher energies.

To examine the absolute accuracy of our computed photon spectra
for more realistic particle spectra, we compared our results
with those of \cite{sturner:97a}.  To facilitate this
comparison, Sturner kindly computed photon spectra for a
particle distribution function of the form
\begin{equation}
N(p) = A'\left(\frac{pc}{\mathrm{1\,MeV}}\right)^{-\Gamma}
\exp\left(-\frac{T}{\cutoff}\right),
\end{equation}
where $A'$ has units of $\cm^{-3}\,({\rm MeV}/c)^{-1}$ and $T$
is the kinetic energy.  Sturner provided synchrotron, inverse
Compton and nonthermal bremsstrahlung spectra for $\Gamma=$
1.8, 2.0 and 2.3 and $\cutoff$=10\,TeV.  The synchrotron
spectrum was computed using $\btot$=1\,$\mu$G. The inverse
Compton spectrum was computed for the cosmic background
radiation field, using a temperature of 2.7\,K. The nonthermal
bremsstrahlung spectrum was computed for a fully ionized target
with ion density 0.11\,cm$^{-3}$ consisting of protons
(0.1\,cm$^{-3}$), alpha-particles (0.01\,cm$^{-3}$), and free
electrons (0.12\,cm$^{-3}$), corresponding to relative weights
of $X_e = 1.090$ and $X_i = 1.182$, for \ee\ and \ez\
bremsstrahlung, respectively. We used our models to compute
photon spectra for the same parameters and particle
distribution function.

Figures \ref{fig:sturner-sync-diff} and
\ref{fig:sturner-gamma-diff} show our spectra divided by those
obtained from Sturner. Over most of the energy range, our
spectra agree quite well with Sturner's; the inverse Compton
and synchrotron spectra agree to within $<1$\%, while the
nonthermal bremsstrahlung spectra typically differ by $\lesssim
5$\%. Aside from the weak $\Gamma$ dependence in the nonthermal
bremsstrahlung differences below about 100\,MeV, the
differences between Sturner's spectra and ours are largely
independent of the spectral index, $\Gamma$. Because the
$\Gamma$ dependent errors in our spectra are constrained by the
recurrence relation, the weak residual $\Gamma$ dependence seen
in the nonthermal bremsstrahlung differences appears to be
associated with Sturner's spectra. The reason for the overall
$\sim$6\% discrepancy below 0.25\,MeV is unclear; features in
the nonthermal bremsstrahlung ratio near 30\,TeV and near
10\,MeV correspond to points in Sturner's spectra at which the
slope changes discontinuously. Because Sturner's synchrotron
and inverse Compton fluxes become negative at the extreme end
of the high-energy photon spectra, rather than asymptotically
approaching zero as ours do, we attribute the larger
differences near the cutoff in these spectra to numerical
errors in Sturner's spectra.  However, since the largest
differences occur at extremely low flux levels, they are
probably not important for observational comparisons.

\begin{figure}[t]
\epsscale{1.1}
\hspace*{-1.5cm}
\plotone{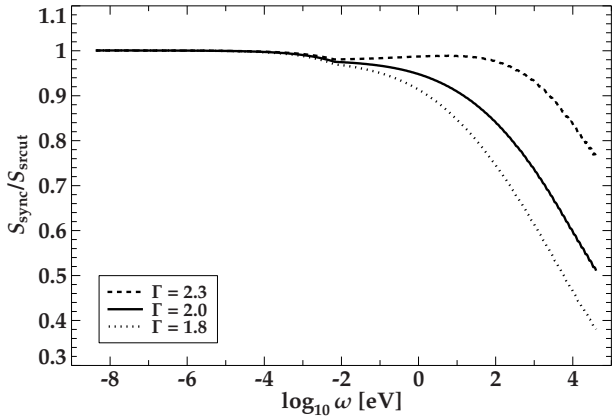}
\caption{\label{fig:srcut}
Ratio showing our synchrotron fluxes divided by \srcut\ synchrotron
fluxes from \xspec\ for $\Gamma=1.8$ (dotted), $\Gamma=2$ (solid),
$\Gamma=2.3$ (dashed). We used $\btot =10\,\mu$G and
$\cutoff=$10\,TeV corresponding to $\nu_\mathrm{break}=1.612
\times 10^{16}$\,Hz.
}
\end{figure}

We also compared our synchrotron model with the \srcut\ model
\citep{reynolds:98a,reynolds:99a} in \xspec\ version {\tt
11.3.2d}. \srcut\ is designed to compute the synchrotron
spectrum from an exponentially cut off power-law distribution
of electrons in a homogeneous magnetic field. It depends on a
break-frequency parameter, $\nu_\mathrm{break}$, defined as the
critical frequency for $\sin\alpha=1$ and $\gamma \equiv
\cutoff / (m_e c^2)$.  For $\nu \ll \nu_\mathrm{break}$, we
verified that \srcut\ is consistent with the analytic result
for a power-law electron distribution. For frequencies
comparable to $\nu_\mathrm{break}$ or larger, we find that
\srcut\ spectra differ from our spectra by as much as a factor
of two or more. Figure \ref{fig:srcut} shows that the ratio of
the two models depends on photon energy and on the power-law
index.  The ratio also depends on the break frequency in the
sense that the models agree for $\nu \ll \nu_\mathrm{break}$
but disagree for frequencies near the break and above.  We are
confident that our computed synchrotron spectrum is accurate,
first, because the validity of equation (\ref{eq:recur})
indicates that any errors are independent of the power-law
index, $\Gamma$, and cutoff-energy, $\cutoff$ and, second,
because our computed spectrum is consistent with Sturner's
model.  We conclude that, for frequencies near and above
$\nu_\mathrm{break}$, the \srcut\ model does not accurately
represent the synchrotron spectrum from an exponentially
cut-off power-law distribution of electrons in a homogeneous
magnetic field.

\begin{figure}[t]
\epsscale{1.1}
\hspace*{-1.5cm}
\plotone{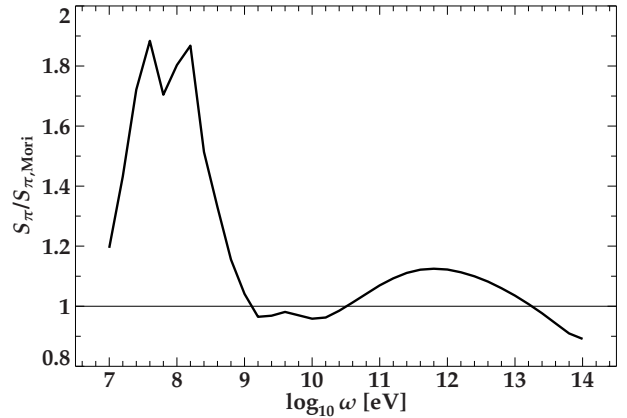}
\caption{\label{fig:mori}
Ratio showing our neutral-pion decay spectrum divided by that of
\citet{mori:97a}.
}
\end{figure}

We tested our neutral-pion decay model by comparing our results
with the results of \cite{mori:97a}. For the purpose of this
comparison, we used his proton momentum distribution function of
the form $N(p) = \left(4\pi/v\right) J_p(p)$ where
\begin{equation}
J_p(p) = 
\begin{cases}
 6.65 \times 10^{-6} \left(\frac{E}{\rm 100\,GeV}\right)^{-2.75},
     \quad E > 100\,\GeV; \\[10pt]
 1.67 \left(\frac{p}{\rm GeV/c}\right)^{-2.7}
      \left[1 + \left(\frac{\rm 2.5\,GeV/c}{p}\right)^2
      \right]^{-1/2}, \\[10pt]
      \hspace*{4cm}
      E \le 100\,\GeV,      
\end{cases}
\end{equation}
where $p \equiv \gamma m_p v$ and $E = T + m_p c^2$. Figure
\ref{fig:mori} shows our computed neutral-pion decay gamma-ray
spectrum divided by the fluxes given in Table 1 of
\cite{mori:97a}.  The largest difference occurs near 100\,MeV,
where our flux is about a factor of two larger than that of
Mori. Because \cite{mori:97a} used a much more detailed model
of pion production, we do not expect exact agreement.  Yet, for
photon energies above 1\,GeV, the two gamma-ray fluxes agree to
within about 10\%.  Because we are primarily interested in
fitting data in the 1-10\,TeV band, our simplified pion-decay
model is adequate for our needs.

\section{Fitting Observed Spectra}
\label{sec:observ}

Fitting a model to an observed spectrum involves minimizing a
goodness-of-fit statistic that compares the observed spectral
data to the spectral model.  This comparison often involves
binned data.

For example, consider X-ray observations which provide the
observed number of counts in each energy bin. Neglecting
nonlinear effects in the X-ray detector, the expected number
of counts is usually computed using an expression of the form
\citep{davis:01a}
\begin{equation}
  C(h) = B(h) + t \int_{\Delta E_h} \diff E\;R(h, E) A(E) S(E),
\label{eq:forward-folding}
\end{equation}
where $C(h)$ is the total number of counts in bin $h$, $B(h)$
is the number of counts due to the instrumental background, $t$
is the exposure time and $S(E)$ is the spectral model
describing the incident flux of photons with energy $E$. In
equation (\ref{eq:forward-folding}), $R(h,E)$ is the redistribution function,
describing the probability that incident photons with energy $E$
contribute counts to bin $h$, and $A(E)$ is the effective
area, accounting for the telescope collecting area, the
transmission efficiency of the optical system and the quantum
efficiency of the detector.

In spectral fitting software, equation (\ref{eq:forward-folding}) is
usually implemented as a discrete sum of the form
\begin{equation}
   C_h = B_h + t \sum_k R_{h,k} A_k S_k.
\end{equation}
The redistribution function, $R(h,E)$, is represented as a
matrix, $R_{h,k}$, and the effective area and source models are
represented as vectors, $A_k$, and $S_k$, respectively.

To accurately represent the integral in equation
(\ref{eq:forward-folding}), the software must compute the
model, $S_k$, as an integral
\begin{equation}
 S_k (E_k; \Delta E_k) \equiv \int_{\Delta E_k}\diff E\;S(E),
 \label{eq:bin-integral}
\end{equation}
over the width $\Delta E_k$ of each spectral bin, $E_k$. For
binned data, we evaluate these integrals using the well-known
Simpson's rule. Although this approach requires three function
evaluations per bin, the bin-edge function values may be shared
between neighboring bins, reducing the total cost of each
spectrum computation by about 30\%. This approach is accurate
to the extent that a quadratic polynomial is a good
approximation to the underlying function, $S(E)$, within each
spectral bin.

\section{Fit Constraints}
\label{sec:constraints}

When simultaneously fitting radio, X-ray and gamma-ray
observations of supernova remnants, the degeneracy of certain
fit parameters (\S\ref{sec:sync}) and the variety of emission
mechanisms in the gamma-ray band may make it difficult to
determine a unique set of fit parameters unless additional
constraints are available.

Based on reasonably general principles, a number of
constraints can sometimes be imposed.  For example:
\begin{enumerate}
 \item When the magnetic field associated with the synchrotron
 emission is generated primarily by cosmic-ray streaming
 \citep{lucek:00a},
 the energy density in cosmic-rays should set an upper limit
 on the magnetic energy density.  The corresponding
 upper limit on the magnetic field strength is
\begin{equation}
   \btot \le \sqrt{8\pi\left(\varepsilon_e + \varepsilon_p\right)},
\end{equation}
where $\varepsilon_e$ and $\varepsilon_p$ are the energy
density in cosmic-ray electrons and protons respectively.
Because $\btot$ and $\cutoff$ are degenerate,
introducing an upper limit on $\btot$ effectively sets a
lower limit on $\cutoff$.
 \item If electrons and protons are injected into the
 accelerator at the same rate, the normalization of the proton
 momentum distribution $A_p$ can be fixed by requiring equal
 densities in nonthermal electrons and protons at
 some characteristic injection kinetic energy
 $T_\mathrm{inj} \ll  m_e c^2$:
 \begin{equation}
       N_e(p_{e,\mathrm{inj}})\diff p_e = N_p(p_{p,\mathrm{inj}})\diff p_p.
 \end{equation}
When both distribution functions are of the form $N(p) =
A(p/p_0)^{-\Gamma}$ and share the same power-law exponent,
$\Gamma$, this constraint implies that $A_p/A_e =
(m_p/m_e)^{(\Gamma-1)/2} \approx 91$ if $\Gamma=2.2$
\citep{bell:78b}. By fixing the ratio $A_p/A_e$, this
constraint reduces the variation in the ratio
$\norm_\pi/\norm_\mathrm{B}$ to the variation associated with
the mass ratio of the associated targets, $n_p V_\pi / (n_0
V_\mathrm{B}$).
\item In many cases it should be reasonable to assume that
the emitting volume that produces synchrotron emission will
also produce inverse Compton emission due to up-scattering of
cosmic background photons.  In such cases, one can impose a
lower limit on the normalization associated with this inverse
Compton process so that $\norm_{I,\mathrm{CBR}} \ge {\cal
N}_\mathrm{S}$. This constraint ensures that fits to the gamma-ray
spectrum will include an appropriate contribution of inverse
Compton emission.  A similar argument can be used to constrain
the minimum value of the nonthermal bremsstrahlung normalization,
$\norm_\mathrm{B}$ for an assumed minimum target density.
\end{enumerate}

To support imposing constraints based on charge conservation,
our software provides a function to compute the proton norm,
$A_p$, that, for a given electron norm, $A_e$, will yield equal
nonthermal electron and proton densities at a given injection
kinetic energy $T_\mathrm{inj}$.  For constraints that depend on
the energy density in nonthermal particles, our software
provides functions that compute the energy density of each
particle population.  The energy density is defined to be
\begin{equation}
  \varepsilon = m c^2 \int_{p_\mathrm{min}}^\infty
  \diff p\;N(p) \left(\gamma(p)-1\right),
\end{equation}
where the lower integration limit, $p_\mathrm{min}$, is the
momentum at which the thermal and nonthermal particle densities
are equal.  Note that $p_\mathrm{min}$ depends on the density
and temperature of the thermal particles and on the nonthermal
particle momentum distribution.

In practice, complicated fit constraints such as the upper
limit on $\btot$ may be imposed using a technique
analogous to the method of Lagrange multipliers
\citep{mathews:65a}.  Rather than minimizing $\chi^2$, the idea
is to construct a constraint function $g({\vect{x}}) \ge 0$
which may depend on a vector of parameters, ${\vect{x}}$, and
then to minimize the sum, $\chi^2 + \lambda g({\vect{x}})$,
where $\lambda$ is a parameter that determines the importance
of the constraint. The constraint function should be
constructed to ensure that $g({\vect{x}}) = 0$ when the
constraint is satisfied.  For example, to impose the constraint
that $\btot^2 < 8\pi(\varepsilon_e + \varepsilon_p)$, one might
choose
\begin{equation}
  g(\vect{x}) = \left\{
     \begin{array}{ll}
       \btot^2 \left[8\pi(\varepsilon_e + \varepsilon_p)\right]^{-1},
          &\btot^2 \ge 8\pi(\varepsilon_e + \varepsilon_p); \\
       0, & {\rm otherwise}.
        \end{array}
        \right.
\end{equation}
Similar terms may be added to impose additional constraints.

\section{Conclusions}
\label{sec:concl}

We have described models that can be used to compute the
synchrotron, inverse Compton, nonthermal bremsstrahlung and
neutral-pion decay spectra of homogeneous sources containing
nonthermal electrons and protons.  We have implemented these
models as an importable module for use in the spectral fitting
package \isis\ \citep{houck:00a}; the software is available from the
\isis\ web page\footnote{\hyperlink{\isisurl}}. The models
are designed to be accurate and fast enough for use in
interactive data analysis on a typical workstation. To the best
of our knowledge, this is the first implementation of some of
these models in a form suitable for interactive fitting with
publicly available data analysis software. Results derived
using the synchrotron and inverse Compton models have been
presented elsewhere by \citet{pannuti:03a} and by
\citet{allen:05a}.

We assessed the accuracy of these models by using a recurrence
relation, by comparing with analytic solutions for synchrotron
and inverse Compton scattering and by comparing with published
work by other authors. The accuracy with which our solutions
obey the recurrence relation in equation
(\ref{eq:recur-finite}) demonstrates their correct dependence
upon the power-law index and cutoff-energy parameters.  We also
showed that, over most of the energy range of interest, our
models agree with those of \citet{sturner:97a}.  The largest
differences between our models and those of
\citet{sturner:97a}, of order $\sim$10\%, occur near the cutoff
of each model spectrum and are essentially independent of the
spectral index.

We found much larger differences between our synchrotron
spectrum and the \srcut\ model from \xspec. Although consistent
with the analytic solution for $\nu \ll \nu_\mathrm{break}$,
\srcut\ differs from our model and from Sturner's by as much as
a factor of two or more in the X-ray band near
$\nu_\mathrm{break}$ (see Figure \ref{fig:srcut}). The most
important difference is that the normalization derived using
\srcut\ overestimates the radio flux at 1\,GHz by an amount
dependent upon the spectral index as shown in Figure
\ref{fig:srcut}. The spectral index from \srcut\ tends to be a
few percent too steep and the break frequency tends to be a few
percent too low but, in practice, such differences may be
detectable only with very high quality data. Note that, by
overestimating the radio flux, \srcut\ may suggest the
existence of positive curvature in the underlying particle
momentum distribution.

In future work, we hope to improve the neutral-pion decay model
by explicitly computing the integral over proton momenta using
improved pion-production cross-sections.  This refinement will
extend the applicable range to sub-GeV photon energies and will
improve the overall accuracy of the model. These improvements
will be more important as better observations of the gamma-ray
spectrum become available from the {\it Gamma-Ray Large Area
Space Telescope} (GLAST)\footnote{\url{http://glast.gsfc.nasa.gov}} 
and from future advances in instrumentation.

\acknowledgments

We thank E. Haug for providing software to compute values for
the \ee\ bremsstrahlung cross-sections and S. Sturner for
providing numerical tables used to verify our results for
selected electron momentum distributions.  We also acknowledge
useful conversations with John E. Davis and a number of
comments from the referee which helped us to improve the paper.

\begin{appendix}

\section{Spectrum Tables}

For reference, Tables \ref{table:sync},
\ref{table:gamma-powerlaw} and \ref{table:gamma-curved} contain
sample spectra for each emission process described in this
paper. The nonthermal electron and proton momentum spectra have the
same shape, with slope $\Gamma = 2$, and a cutoff energy
$\cutoff = 10$\,TeV. Table \ref{table:sync} gives
synchrotron spectra for curvatures $a=0$ and $a=0.05$. Tables
\ref{table:gamma-powerlaw} and \ref{table:gamma-curved} give
the inverse Compton, nonthermal bremsstrahlung and neutral-pion
decay spectra for curvatures $a=0$ and $a=0.05$,
respectively. All normalization coefficients were set to unity.
The synchrotron spectrum was computed for $\btot =
10\,\mu$G. The inverse Compton spectrum was computed
using a 2.725\,K blackbody distribution of seed photons
\citep{bennett:03a}.  Contributions to nonthermal
bremsstrahlung due to the \ee\ and \ep\ processes are listed
separately, with their respective weights set to unity.

\section{Distributed Computation}

The spectral models described in this paper were designed to
fit multi-wavelength spectral data interactively on a typical
workstation and to achieve a high degree of accuracy as
efficiently as possible. In practice, to be sure that the fit
has fully converged and to derive confidence limits, it is
necessary to thoroughly examine the parameter space in the
neighborhood of the best-fit parameters.  To conduct this
search more quickly, we have found it useful to distribute the
task of computing single-parameter confidence limits over a
number processors running in parallel \citep{noble:05a}. One
master process manages the computations being performed by a
number of slave processes that run on different computers.  All
of the computers are connected by a local network and all
processes have access to the relevant data and spectral models.
The master process assigns each slave process the task of
computing confidence limits for a single parameter.  If a slave
process finds an improved fit, that new parameter set is sent to
the master process.  If that fit is the best yet found by
any slave, the master commands all the slave processes to
re-start the search from the beginning, using the new
parameter set.  We have found that the time required to obtain
a set of converged single parameter confidence limits using this
approach is often reduced by more than a factor of $N$, where
$N$ is the number of slave processes.

We have implemented this algorithm using the Parallel Virtual
Machine (\pvm) \citep{geist:94a} to handle message passing
between the master and slave processes.  We used the spectral
fitting package, \isis\, \citep{houck:00a} to perform model
fits and confidence limit searches using a set of \slang\
scripts. An \slang\ module provides a scriptable interface to
the \pvm\ library, making it possible for these scripts to
communicate with the \pvm.

\clearpage


\begin{deluxetable}{cccccc}
\tabletypesize{\scriptsize}
\tablecaption{Sample Synchrotron Spectra\label{table:sync}}
\tablewidth{0pt}
\tablehead{
    \colhead{}
    & \multicolumn{2}{c}{Flux}
    & \colhead{}
    & \multicolumn{2}{c}{Flux} \\
    \colhead{Energy}
    & \colhead{a=0}
    & \colhead{a=0.05}
    & \colhead{Energy}
    & \colhead{a=0}
    & \colhead{a=0.05} \\
    \colhead{(eV)}
    & \multicolumn{2}{c}{(photons~s$^{-1}$~cm$^{-2}$~GeV$^{-1}$)}
    & \colhead{(eV)}
    & \multicolumn{2}{c}{(photons~s$^{-1}$~cm$^{-2}$~GeV$^{-1}$)}
}
\startdata
{\tt 1.0000e-07} & {\tt 6.8294e+15} & {\tt 6.8613e+15} & {\tt 1.7783e-01} & {\tt 2.5564e+06} & {\tt 7.0374e+06}\\
{\tt 1.7783e-07} & {\tt 2.8798e+15} & {\tt 2.9011e+15} & {\tt 3.1623e-01} & {\tt 1.0403e+06} & {\tt 3.1004e+06}\\
{\tt 3.1623e-07} & {\tt 1.2144e+15} & {\tt 1.2284e+15} & {\tt 5.6234e-01} & {\tt 4.1918e+05} & {\tt 1.3546e+06}\\
{\tt 5.6234e-07} & {\tt 5.1206e+14} & {\tt 5.2114e+14} & {\tt 1.0000e+00} & {\tt 1.6680e+05} & {\tt 5.8530e+05}\\
{\tt 1.0000e-06} & {\tt 2.1591e+14} & {\tt 2.2168e+14} & {\tt 1.7783e+00} & {\tt 6.5346e+04} & {\tt 2.4928e+05}\\
{\tt 1.7783e-06} & {\tt 9.1040e+13} & {\tt 9.4594e+13} & {\tt 3.1623e+00} & {\tt 2.5107e+04} & {\tt 1.0423e+05}\\
{\tt 3.1623e-06} & {\tt 3.8386e+13} & {\tt 4.0507e+13} & {\tt 5.6234e+00} & {\tt 9.4173e+03} & {\tt 4.2578e+04}\\
{\tt 5.6234e-06} & {\tt 1.6184e+13} & {\tt 1.7409e+13} & {\tt 1.0000e+01} & {\tt 3.4291e+03} & {\tt 1.6897e+04}\\
{\tt 1.0000e-05} & {\tt 6.8229e+12} & {\tt 7.5091e+12} & {\tt 1.7783e+01} & {\tt 1.2040e+03} & {\tt 6.4693e+03}\\
{\tt 1.7783e-05} & {\tt 2.8762e+12} & {\tt 3.2506e+12} & {\tt 3.1623e+01} & {\tt 4.0433e+02} & {\tt 2.3701e+03}\\
{\tt 3.1623e-05} & {\tt 1.2123e+12} & {\tt 1.4121e+12} & {\tt 5.6234e+01} & {\tt 1.2859e+02} & {\tt 8.2272e+02}\\
{\tt 5.6234e-05} & {\tt 5.1092e+11} & {\tt 6.1560e+11} & {\tt 1.0000e+02} & {\tt 3.8275e+01} & {\tt 2.6741e+02}\\
{\tt 1.0000e-04} & {\tt 2.1528e+11} & {\tt 2.6929e+11} & {\tt 1.7783e+02} & {\tt 1.0511e+01} & {\tt 8.0233e+01}\\
{\tt 1.7783e-04} & {\tt 9.0683e+10} & {\tt 1.1819e+11} & {\tt 3.1623e+02} & {\tt 2.6180e+00} & {\tt 2.1844e+01}\\
{\tt 3.1623e-04} & {\tt 3.8186e+10} & {\tt 5.2042e+10} & {\tt 5.6234e+02} & {\tt 5.7921e-01} & {\tt 5.2861e+00}\\
{\tt 5.6234e-04} & {\tt 1.6072e+10} & {\tt 2.2985e+10} & {\tt 1.0000e+03} & {\tt 1.1101e-01} & {\tt 1.1090e+00}\\
{\tt 1.0000e-03} & {\tt 6.7607e+09} & {\tt 1.0181e+10} & {\tt 1.7783e+03} & {\tt 1.7883e-02} & {\tt 1.9570e-01}\\
{\tt 1.7783e-03} & {\tt 2.8416e+09} & {\tt 4.5213e+09} & {\tt 3.1623e+03} & {\tt 2.3343e-03} & {\tt 2.8009e-02}\\
{\tt 3.1623e-03} & {\tt 1.1931e+09} & {\tt 2.0124e+09} & {\tt 5.6234e+03} & {\tt 2.3621e-04} & {\tt 3.1107e-03}\\
{\tt 5.6234e-03} & {\tt 5.0025e+08} & {\tt 8.9740e+08} & {\tt 1.0000e+04} & {\tt 1.7564e-05} & {\tt 2.5414e-04}\\
{\tt 1.0000e-02} & {\tt 2.0938e+08} & {\tt 4.0070e+08} & {\tt 1.7783e+04} & {\tt 8.9943e-07} & {\tt 1.4316e-05}\\
{\tt 1.7783e-02} & {\tt 8.7436e+07} & {\tt 1.7903e+08} & {\tt 3.1623e+04} & {\tt 2.9327e-08} & {\tt 5.1409e-07}\\
{\tt 3.1623e-02} & {\tt 3.6404e+07} & {\tt 7.9973e+07} & {\tt 5.6234e+04} & {\tt 5.5363e-10} & {\tt 1.0702e-08}\\
{\tt 5.6234e-02} & {\tt 1.5099e+07} & {\tt 3.5680e+07} & {\tt 1.0000e+05} & {\tt 5.3930e-12} & {\tt 1.1512e-10}\\
{\tt 1.0000e-01} & {\tt 6.2322e+06} & {\tt 1.5878e+07} & {\tt 1.7783e+05} & {\tt 2.3578e-14} & {\tt 5.5658e-13}\\
\enddata
\end{deluxetable}

\begin{deluxetable}{lrrrr}
\tabletypesize{\scriptsize}
\tablecaption{Sample Gamma-Ray Spectra ($a=0$)\label{table:gamma-powerlaw}}
\tablewidth{0pt}
\tablehead{%
    \colhead{}
    & \multicolumn{4}{c}{Flux} \\
    \colhead{Energy}
    & \colhead{Inv. Comp.}
    & \colhead{$ee$ Brem.}
    & \colhead{$ep$ Brem.}
    & \colhead{$\pi^0$ decay} \\
    \colhead{(eV)}
    & \multicolumn{4}{c}{(photons~s$^{-1}$~cm$^{-2}$~GeV$^{-1}$)}
}
\startdata
{\tt 1.0000e+06} & {\tt 2.0062e-10} & {\tt 1.3845e-10} & {\tt 1.6562e-10} & {\tt 8.2454e-18}\\
{\tt 1.7783e+06} & {\tt 8.4541e-11} & {\tt 5.3800e-11} & {\tt 6.2697e-11} & {\tt 2.4513e-17}\\
{\tt 3.1623e+06} & {\tt 3.5616e-11} & {\tt 2.0632e-11} & {\tt 2.3393e-11} & {\tt 7.0967e-17}\\
{\tt 5.6234e+06} & {\tt 1.5000e-11} & {\tt 7.7773e-12} & {\tt 8.5860e-12} & {\tt 1.9707e-16}\\
{\tt 1.0000e+07} & {\tt 6.3147e-12} & {\tt 2.8778e-12} & {\tt 3.1021e-12} & {\tt 5.1353e-16}\\
{\tt 1.7783e+07} & {\tt 2.6568e-12} & {\tt 1.0460e-12} & {\tt 1.1053e-12} & {\tt 1.2201e-15}\\
{\tt 3.1623e+07} & {\tt 1.1170e-12} & {\tt 3.7410e-13} & {\tt 3.8916e-13} & {\tt 2.5548e-15}\\
{\tt 5.6234e+07} & {\tt 4.6913e-13} & {\tt 1.3195e-13} & {\tt 1.3565e-13} & {\tt 2.7122e-15}\\
{\tt 1.0000e+08} & {\tt 1.9677e-13} & {\tt 4.6000e-14} & {\tt 4.6881e-14} & {\tt 2.7122e-15}\\
{\tt 1.7783e+08} & {\tt 8.2389e-14} & {\tt 1.5880e-14} & {\tt 1.6085e-14} & {\tt 1.9824e-15}\\
{\tt 3.1623e+08} & {\tt 3.4416e-14} & {\tt 5.4375e-15} & {\tt 5.4842e-15} & {\tt 9.0019e-16}\\
{\tt 5.6234e+08} & {\tt 1.4332e-14} & {\tt 1.8492e-15} & {\tt 1.8596e-15} & {\tt 3.6492e-16}\\
{\tt 1.0000e+09} & {\tt 5.9441e-15} & {\tt 6.2516e-16} & {\tt 6.2745e-16} & {\tt 1.3641e-16}\\
{\tt 1.7783e+09} & {\tt 2.4521e-15} & {\tt 2.1026e-16} & {\tt 2.1075e-16} & {\tt 4.8273e-17}\\
{\tt 3.1623e+09} & {\tt 1.0045e-15} & {\tt 7.0378e-17} & {\tt 7.0481e-17} & {\tt 1.6490e-17}\\
{\tt 5.6234e+09} & {\tt 4.0771e-16} & {\tt 2.3445e-17} & {\tt 2.3468e-17} & {\tt 5.5070e-18}\\
{\tt 1.0000e+10} & {\tt 1.6351e-16} & {\tt 7.7725e-18} & {\tt 7.7775e-18} & {\tt 1.8116e-18}\\
{\tt 1.7783e+10} & {\tt 6.4567e-17} & {\tt 2.5626e-18} & {\tt 2.5637e-18} & {\tt 5.8906e-19}\\
{\tt 3.1623e+10} & {\tt 2.4988e-17} & {\tt 8.3928e-19} & {\tt 8.3953e-19} & {\tt 1.8934e-19}\\
{\tt 5.6234e+10} & {\tt 9.4223e-18} & {\tt 2.7251e-19} & {\tt 2.7257e-19} & {\tt 5.9973e-20}\\
{\tt 1.0000e+11} & {\tt 3.4360e-18} & {\tt 8.7461e-20} & {\tt 8.7476e-20} & {\tt 1.8582e-20}\\
{\tt 1.7783e+11} & {\tt 1.2001e-18} & {\tt 2.7620e-20} & {\tt 2.7624e-20} & {\tt 5.5578e-21}\\
{\tt 3.1623e+11} & {\tt 3.9651e-19} & {\tt 8.5250e-21} & {\tt 8.5262e-21} & {\tt 1.5693e-21}\\
{\tt 5.6234e+11} & {\tt 1.2191e-19} & {\tt 2.5465e-21} & {\tt 2.5469e-21} & {\tt 4.0316e-22}\\
{\tt 1.0000e+12} & {\tt 3.4130e-20} & {\tt 7.2548e-22} & {\tt 7.2560e-22} & {\tt 8.8745e-23}\\
{\tt 1.7783e+12} & {\tt 8.4456e-21} & {\tt 1.9286e-22} & {\tt 1.9290e-22} & {\tt 1.5161e-23}\\
{\tt 3.1623e+12} & {\tt 1.7719e-21} & {\tt 4.6256e-23} & {\tt 4.6267e-23} & {\tt 1.7037e-24}\\
{\tt 5.6234e+12} & {\tt 2.9674e-22} & {\tt 9.4868e-24} & {\tt 9.4896e-24} & {\tt 9.4934e-26}\\
{\tt 1.0000e+13} & {\tt 3.6207e-23} & {\tt 1.5226e-24} & {\tt 1.5232e-24} & {\tt 1.6037e-27}\\
{\tt 1.7783e+13} & {\tt 2.7845e-24} & {\tt 1.6433e-25} & {\tt 1.6443e-25} & {\tt 3.4422e-30}\\
{\tt 3.1623e+13} & {\tt 1.0582e-25} & {\tt 9.1509e-27} & {\tt 9.1588e-27} & {\tt 1.9973e-34}\\
{\tt 5.6234e+13} & {\tt 1.2918e-27} & {\tt 1.6454e-28} & {\tt 1.6477e-28} & {\tt 1.9824e-41}\\
{\tt 1.0000e+14} & {\tt 2.3039e-30} & {\tt 4.1544e-31} & {\tt 4.1640e-31} & {\tt 2.4571e-53}\\
\enddata
\end{deluxetable}

\begin{deluxetable}{lrrrrrr}
\tabletypesize{\scriptsize}
\tablecaption{Sample Gamma-Ray Spectra ($a=0.05$)\label{table:gamma-curved}}
\tablewidth{0pt}
\tablehead{
    \colhead{}
    & \multicolumn{4}{c}{Flux} \\
    \colhead{Energy}
    & \colhead{Inv. Comp.}
    & \colhead{$ee$ Brem.}
    & \colhead{$ep$ Brem.}
    & \colhead{$\pi^0$ decay} \\
    \colhead{(eV)}
    & \multicolumn{4}{c}{(photons~s$^{-1}$~cm$^{-2}$~GeV$^{-1}$)}
}
\startdata
{\tt 1.0000e+06} & {\tt 2.4116e-10} & {\tt 1.3856e-10} & {\tt 1.6572e-10} & {\tt 1.1455e-17}\\
{\tt 1.7783e+06} & {\tt 1.0553e-10} & {\tt 5.3856e-11} & {\tt 6.2750e-11} & {\tt 3.1243e-17}\\
{\tt 3.1623e+06} & {\tt 4.6340e-11} & {\tt 2.0662e-11} & {\tt 2.3422e-11} & {\tt 8.4245e-17}\\
{\tt 5.6234e+06} & {\tt 2.0415e-11} & {\tt 7.7932e-12} & {\tt 8.6015e-12} & {\tt 2.2119e-16}\\
{\tt 1.0000e+07} & {\tt 9.0225e-12} & {\tt 2.8864e-12} & {\tt 3.1106e-12} & {\tt 5.5285e-16}\\
{\tt 1.7783e+07} & {\tt 3.9996e-12} & {\tt 1.0506e-12} & {\tt 1.1099e-12} & {\tt 1.2763e-15}\\
{\tt 3.1623e+07} & {\tt 1.7779e-12} & {\tt 3.7663e-13} & {\tt 3.9168e-13} & {\tt 2.6248e-15}\\
{\tt 5.6234e+07} & {\tt 7.9236e-13} & {\tt 1.3332e-13} & {\tt 1.3701e-13} & {\tt 2.7832e-15}\\
{\tt 1.0000e+08} & {\tt 3.5389e-13} & {\tt 4.6739e-14} & {\tt 4.7619e-14} & {\tt 2.7832e-15}\\
{\tt 1.7783e+08} & {\tt 1.5832e-13} & {\tt 1.6278e-14} & {\tt 1.6483e-14} & {\tt 2.0479e-15}\\
{\tt 3.1623e+08} & {\tt 7.0899e-14} & {\tt 5.6505e-15} & {\tt 5.6971e-15} & {\tt 9.5028e-16}\\
{\tt 5.6234e+08} & {\tt 3.1756e-14} & {\tt 1.9621e-15} & {\tt 1.9725e-15} & {\tt 3.9827e-16}\\
{\tt 1.0000e+09} & {\tt 1.4211e-14} & {\tt 6.8424e-16} & {\tt 6.8653e-16} & {\tt 1.5601e-16}\\
{\tt 1.7783e+09} & {\tt 6.3444e-15} & {\tt 2.4051e-16} & {\tt 2.4101e-16} & {\tt 5.8712e-17}\\
{\tt 3.1623e+09} & {\tt 2.8206e-15} & {\tt 8.5278e-17} & {\tt 8.5385e-17} & {\tt 2.1653e-17}\\
{\tt 5.6234e+09} & {\tt 1.2459e-15} & {\tt 3.0488e-17} & {\tt 3.0513e-17} & {\tt 7.9260e-18}\\
{\tt 1.0000e+10} & {\tt 5.4508e-16} & {\tt 1.0983e-17} & {\tt 1.0988e-17} & {\tt 2.9006e-18}\\
{\tt 1.7783e+10} & {\tt 2.3532e-16} & {\tt 3.9805e-18} & {\tt 3.9818e-18} & {\tt 1.0644e-18}\\
{\tt 3.1623e+10} & {\tt 9.9771e-17} & {\tt 1.4483e-18} & {\tt 1.4486e-18} & {\tt 3.9144e-19}\\
{\tt 5.6234e+10} & {\tt 4.1292e-17} & {\tt 5.2743e-19} & {\tt 5.2752e-19} & {\tt 1.4370e-19}\\
{\tt 1.0000e+11} & {\tt 1.6555e-17} & {\tt 1.9143e-19} & {\tt 1.9145e-19} & {\tt 5.2220e-20}\\
{\tt 1.7783e+11} & {\tt 6.3677e-18} & {\tt 6.8852e-20} & {\tt 6.8859e-20} & {\tt 1.8516e-20}\\
{\tt 3.1623e+11} & {\tt 2.3209e-18} & {\tt 2.4353e-20} & {\tt 2.4355e-20} & {\tt 6.2578e-21}\\
{\tt 5.6234e+11} & {\tt 7.8871e-19} & {\tt 8.3813e-21} & {\tt 8.3822e-21} & {\tt 1.9414e-21}\\
{\tt 1.0000e+12} & {\tt 2.4462e-19} & {\tt 2.7658e-21} & {\tt 2.7661e-21} & {\tt 5.2052e-22}\\
{\tt 1.7783e+12} & {\tt 6.7266e-20} & {\tt 8.5667e-22} & {\tt 8.5679e-22} & {\tt 1.0934e-22}\\
{\tt 3.1623e+12} & {\tt 1.5746e-20} & {\tt 2.4116e-22} & {\tt 2.4121e-22} & {\tt 1.5278e-23}\\
{\tt 5.6234e+12} & {\tt 2.9588e-21} & {\tt 5.8625e-23} & {\tt 5.8639e-23} & {\tt 1.0730e-24}\\
{\tt 1.0000e+13} & {\tt 4.0826e-22} & {\tt 1.1297e-23} & {\tt 1.1301e-23} & {\tt 2.3213e-26}\\
{\tt 1.7783e+13} & {\tt 3.5900e-23} & {\tt 1.4881e-24} & {\tt 1.4888e-24} & {\tt 6.4948e-29}\\
{\tt 3.1623e+13} & {\tt 1.5844e-24} & {\tt 1.0307e-25} & {\tt 1.0315e-25} & {\tt 5.0034e-33}\\
{\tt 5.6234e+13} & {\tt 2.2946e-26} & {\tt 2.3534e-27} & {\tt 2.3565e-27} & {\tt 6.7148e-40}\\
{\tt 1.0000e+14} & {\tt 4.9900e-29} & {\tt 7.7046e-30} & {\tt 7.7217e-30} & {\tt 1.1451e-51}\\
\enddata
\end{deluxetable}

\end{appendix}

\clearpage


\vspace*{1cm}

\noindent{\it Note added in proof.---}After this manuscript was
submitted, papers by Kamae {\it et al.} (astro-ph/0605581) and
by Kelner, Aharonian \& Bugayov (astro-ph/0606058) appeared,
presenting parameterized cross-sections for various particles
produced of $pp$ collisions. We plan to revise our pion-decay
model to incorporate these new cross-sections.  Details will be
discussed in a subsequent paper.

\end{document}